\documentclass[prb,aps,twocolumn,showpacs]{revtex4}
\usepackage{graphicx,latexsym}
\usepackage{dcolumn}
\usepackage{amsmath,amssymb,epsf,bm}

\begin{document}


\title{Spin orientation and Spin-Hall effect induced by tunneling electrons}
\author{A.G. Mal'shukov$^{1,2,3}$ and C.S.Chu$^{2}$}
\affiliation{$^1$Institute of Spectroscopy, Russian Academy of
Science, 142190 Troitsk, Moscow oblast, Russia \\
$^2$Department of Electrophysics, National Chiao Tung university, Hsinchu 30010, Taiwan \\
$^3$National Center for Theoretical Sciences, Physics Division,
Hsinchu 30043, Taiwan}
\begin{abstract}
It is shown that a flux of unpolarized electrons across a
symmetric double barrier quantum well induces a spin polarization
inside the well. Besides, the transmitted current acquires a spin
polarized component and the spin-Hall current flows in the planar
direction. These phenomena are due to a combined effect of
Dresselhaus interaction and the spin-orbit interaction induced by
gradients of heterostructure material parameters. In contrast to
previous studies of the spin filtering effect, we predict that it
can be observed in case of an isotropic distribution of incident
electrons.
\end{abstract}
\pacs{71.70.Ej, 72.25.Dc, 73.40.Lq}

\maketitle

\section{Introduction}
The spin-orbit interaction (SOI) is a fundamental quantum
relativistic phenomenon which recently attracted much interest in
connection with spin transport of electrons in semiconductors and
metals. Due to SOI, an electric field can influence the spin
degree of freedom, thereby giving rise to a number of transport
phenomena which have potential for application in spintronics. One
of them is the spin-Hall effect (SHE), which recently has been
intensively studied both theoretically (for a review see
\onlinecite{theor}) and experimentally. \cite{exp,Awschalom} A
standard system to study this effect is a 2D electron gas confined
within a quantum well. Due to SHE the electric current in the
quantum well gives rise to a perpendicular flux of the spin
polarization, as well as to the out of plane spin density near
sample edges. Closely related to SHE is the electric spin
orientation, which is a bulk inplane  spin polarization induced by
the DC electric current. \cite{Edelstein} In both cases SOI is
realized either via impurity scattering, or due to an intrinsic
spin-orbit coupling mechanism. The latter consists of two parts.
The first one is the Dresselhaus interaction \cite{Dresselhaus}
which is inherent to all zinc-blende semiconductors. The second
contribution is determined by gradients of material parameters and
the electric potential across a heterointerface.
\cite{Bassani,Winkler} In quantum wells (QW) these interactions
are averaged with wave functions of confinement coordinate in the
heterostructure growth direction. After such an averaging
procedure the second term transforms into the Rashba SOI
\cite{RashbaSOI} which is not zero only in asymmetric in
$z$-direction (growth direction) heterostructures. The Rashba and
averaged Dresselhaus interactions are basic SOI widely used in
works on SHE, as well as in works on spin-dependent transport in
general.

In this work we will consider spin-orbit effects on electron
transport from a different point of view. Namely, we will consider
the electric current parallel to the growth direction, rather than
parallel to the quantum well. An appropriate model for studying such
a situation is a double barrier quantum well. A principal difference
from the conventional 2D system is that one can not use neither
Rashba nor Dresselhaus interactions averaged over QW confinement.
For example, a part of the latter interaction, which is proportional
to the $z$-component of the electron momentum operator, turns to 0
when averaged with a wave function of a confined state in the well.
At the same time it is finite for tunneling states. Considering
these states it is also easy to see that SOI associated with
electric fields at heterointerfaces does not reduce to the single
parameter Rashba interaction. The explicit dependence on $z$ of the
initial spin-orbit interaction becomes important. To make this point
more clear we deliberately considered a symmetric heterostructure,
where Rashba SOI is zero. For such a model of a symmetric double
barrier quantum well we found out that the electric current in the
$z$-direction induces a parallel to the $z$-axis spin density, as
well as a current of spins polarized in a planar direction. The spin
current flows parallel to heterointerfaces. The former effect is an
analog of the electric spin orientation, while the latter is the
spin-Hall effect. Besides, we found the transmittance of the double
barrier structure to be dependent on the spin orientation. A similar
spin filtering effect has been considered before within various
models. \cite{Voskoboynikov, Perel} It was shown there that this
effect can be observed only in case of an anisotropic distribution
of incident electrons. It was suggested to create such a
distribution applying a planar electric field. In contrast to these
works, we predict the spin filtering effect for an isotropic
distribution of electrons tunneling through a symmetric double
barrier structure. Moreover, our spin filter makes electrons to be
polarized in z-direction, instead of the planar polarization in Ref.
\onlinecite{Voskoboynikov, Perel}. Such fundamental distinctions
arise from the $\sim k_z$ term of the Dresselhaus interaction which
has been neglected in Ref. \onlinecite{Perel}.

The general Hamiltonian of the problem will be derived in Sec.II.
In Sec. III we will present our results related to the spin-Hall
effect, spin orientation and the spin filtering effect. A brief
conclusion is presented in Sec. IV.

\section{Spin-orbit Hamiltonian}

Let us consider a quantum well (QW) of the width $2d$ separated
from the left ($z<-d-b$) and right ($z>d+b$) parts of a doped
semiconductor system by two equal barriers of the thickness $b$.
These parts are assumed to be thermal reservoirs with respective
chemical potentials $\mu_l$ and $\mu_r$. Besides a system
homogeneous in $x,y$ directions, we will also consider an electron
gas confined in the $y$-direction. For simplicity, the adiabatic
case will be considered when the confinement width $w$ slowly
increases from QW towards reservoirs. This situation is realized
when the confinement is achieved by depleting the electron gas
with the help of electrodes on top of QW.

The spin-orbit interaction in such a system is represented by two
Hamiltonians $H^{(1)}_{so}$ and $H^{(2)}_{so}$. The former is the
Dresselhaus SOI, while the latter is SOI due to change of the band
gap width and other material parameters across heterointerfaces.
Usually, in narrow gap semiconductors $H^{(2)}_{so}$ is much
stronger. Therefore, we will consider the Dresselhaus interaction
within the first order perturbation theory. Only the part of
$H^{(1)}_{so}$ which is proportional to the $z$-component of the
electron momentum operator will be taken into account. Hence, for
[001] growth direction
\begin{equation}\label{H1}
H^{(1)}_{so}= -i \gamma  \sigma_z (k^2_x -
k^2_y)\frac{\partial}{\partial z}\,,
\end{equation}
where $\sigma_z$ is the Pauli matrix and $\gamma$ is the coupling
parameter, which is assumed to be $z$-independent. An important
property of Hamiltonian (\ref{H1}) is that its expectation values
taken with tunnelling states incident from the left and from the
right reservoirs have opposite signs, while it does not change sign
when $k_x, k_y \rightarrow -k_x, -k_y$. Alternatively, other parts
of the Dresselhaus SOI, which have been omitted in (\ref{H1}), are
even functions of $\hat{k}_z$ and odd functions of $k_x, k_y$.
Since, the interaction represented by $H^{(2)}_{so}$ is of the same
symmetry, the omitted terms of $H^{(1)}_{so}$ do not add
particularly new qualitative features to spin dependent electron
tunneling, as well as to other effects considered below. At the same
time, the symmetry difference of (\ref{H1}) and $H^{(2)}_{so}$ has
important consequences for these effects. This is also the main
reason why the results of Ref. \onlinecite{Perel}, where interaction
(\ref{H1}) has been neglected, are qualitatively different from
those presented below.

Following Ref. \onlinecite{Bassani} the Hamiltonian $H^{(2)}_{so}$
can be written as
\begin{equation}\label{H2}
H^{(2)}_{so}=\frac{1}{k_{\|}}\left(\sigma_x k_y -\sigma_y
k_x\right)h(z)\,,
\end{equation}
where
\begin{equation}\label{h}
h(z)=k_{\|}\frac{\partial\beta}{\partial z}\,
\end{equation}
and $k_{\|}=\sqrt{k_x^2+k_y^2}$. The parameter $h(z)$ denotes SOI
strength which varies across the heterostructure depending on
semiconductor material parameters and the electric potential.
Ignoring the electron energy, which is much less than the gap
value, $\beta(z)$  can be written as \cite{Bassani}
\begin{equation}\label{beta}
\beta(z)=\frac{1}{2m(z)}\frac{\Delta(z)}{3E_{g}(z)+\Delta(z)}\,,
\end{equation}
where $E_{g}(z)$ and $\Delta(z)$ are respective values of the band
gaps and split off energies.

The major effect of SOI (\ref{H1}) is that it gives rise to spin
precession around the $z$-axis. This precession takes place during
particle transmission through the double barrier structure. Since
the width of this structure is small in comparison with the spin
precession length, the effect of the spin precession is expected
to be small. In order to get explicitly the corresponding small
parameter, the Hamiltonian can be transformed using an appropriate
unitary transformation. Taking into account that the kinetic
energy operator in $z$-direction is
\begin{equation}\label{kin}
\frac{1}{2}\frac{\partial}{\partial
z}\frac{1}{m(z)}\frac{\partial}{\partial z}
\end{equation}
one can apply the unitary transformation
\begin{equation}\label{Huni}
H \rightarrow U^{-1}HU \,,
\end{equation}
with
\begin{equation}\label{uni}
U=e^{i\sigma_z \vartheta(z)}\,,
\end{equation}
where
\begin{equation}\label{theta}
\vartheta(z)=-\gamma (k^2_x - k^2_y)\int_0^z m(z)dz\,.
\end{equation}
This transformation removes (\ref{H1}) from the Hamiltonian. At
the same time, applying it to (\ref{H2}) one obtains, up to the
linear in $\vartheta(z)$ terms, the spin orbit interaction
\begin{equation}\label{Hso}
H_{so}=H^{(2)}_{so} + \frac{2\vartheta(z)}{k_{\|}}\left(\sigma_x k_x
+\sigma_y k_y\right)h(z)\,.
\end{equation}
For the model under consideration, with rectangular symmetric
barriers and a rectangular QW, $h(z)$ becomes
\begin{eqnarray}\label{h2}
h(z)&=&k_{\|}[(\beta_r-\beta_b)\left(\delta(z-b-d)-\delta(z+b+d)\right)
\nonumber
\\ &+&(\beta_b-\beta_w)\left(\delta(z-d)-\delta(z+d)\right)]\,.
\end{eqnarray}
The parameters $\beta_r$, $\beta_b$ and $\beta_w$ denote SOI
strengths for reservoirs, barriers and QW, respectively.

The transmission wave functions are represented by two sets of
functions incident from the left ($\psi^l$) and from the right
($\psi^r$) of the double barrier structure. In the zeroth order,
when the second term in (\ref{Hso}) is ignored, these functions
can be conveniently written using the chiral basis. In case of the
quantum wire confinement this basis corresponds to the spin
quantization axis directed along the $y$-axis. Outside the double
barrier structure the scattering eigenstates are represented by
incident, transmitted and reflected plane waves, with transmission
and reflection amplitudes $t^{l/r}_{\sigma}$ and
$r^{l/r}_{\sigma}$, respectively, where $\sigma=1,2$ denotes the
spin projection in the corresponding chiral basis. This projection
is conserved upon the scattering, as far as the second term in
(\ref{Hso}) is neglected. The wave vector of the scattering states
is denoted as $k=\sqrt{2m_r(E-E_{\|})}$, where $m_r$ is the
electron effective mass in the left and right reservoirs, $E$ is
the total energy and $E_{\|}$ is the energy of motion in $x,y$
directions.

\section{Spin current and spin polarization}

\subsection{Spin current}

The non zero spin current $J^s_n$, where $s$ and $n$ denote the
spin polarization and current direction, respectively, can be
calculated already in the zeroth order with respect to
$\vartheta(z)$, while the latter will be important below in
calculation of the spin density and spin dependent transmission.
Hence, in this subsection we entirely neglect the presumably weak
Dresselhaus SOI. Let us take the chiral component of $J^s_n$. That
means that we look for the flux of spins polarized perpendicular
to the flux direction, that can be expressed as
$J^s_n=J\varepsilon^{snz}$, where $\varepsilon^{snz}$ is the
antisymmetric tensor. Using the conventional definition of the
spin current operator $\hat{J}^s_n=\{v_n,\sigma_s/4\}$, where the
spin dependent part of
\begin{equation}\label{v}
v_n=\frac{k_n}{m(z)}+\varepsilon^{niz}\sigma_i \frac{h(z)}{k_{\|}}
\end{equation}
is obtained from (\ref{H2}) and (\ref{h}), the spin current
density can be written as
\begin{eqnarray}\label{J}
J^s_n(z)&=&\frac{\varepsilon^{snz}}{2}\sum_{\mathbf{k}_{\|},\gamma}\int_0^\infty
\frac{dk}{2\pi}[\frac{k_{\|}}{2m^*(z)}(|\psi_1^{\gamma}|^2-|\psi_2^{\gamma}|^2)
\nonumber \\
&+&
\frac{h(z)}{k_{\|}}(|\psi_1^{\gamma}|^2+|\psi_2^{\gamma}|^2)]n_F^{\gamma}(E)
\,,
\end{eqnarray}
where $\gamma=l,r$ and $n_F^{\gamma}(E)$ is the Fermi distribution
function for the left and right reservoirs. The first term in
square brackets represents the bulk spin current density
distributed in QW, barriers and outside, while the second term is
the "surface" term which, according to (\ref{h2}), is finite only
on heterostructure interfaces. From Eq.(\ref{J}) it becomes
immediately evident that the spin current is not zero in the
equilibrium state when $n_F^{l}(E)=n_F^{r}(E)$. For example, the
surface current at each interface is given by
$\rho(z)\Delta\beta(z)$, where $\rho(z)$ is the local equilibrium
electron density and $\Delta\beta(z)$ is the difference of the
spin-orbit coupling parameters $\beta$ on both sides of the
interface, as follows from (\ref{h2}). In a symmetric QW the
surface currents on opposite interfaces flow in opposite
directions, so that they cancel each other. It is easy to see that
the total equilibrium current, obtained by integration of Eq.
(\ref{J}) over $z$, is identically zero in case of a symmetric
heterostructure. That follows from the symmetry relation
$\psi^{l/r}_1(z)=\psi^{r/l}_2(-z)$ which makes $J^s_n(z)$ to be an
odd function of $z$. At the same time, one can not expect the
total current to be zero in an asymmetric heterostructure, as has
been shown by Rashba \cite{Rashba} for confined states. It should
be noted that Rashba found that the total equilibrium current is
cubic with respect to the spin-orbit coupling constant, while the
current density given by Eq. (\ref{J}) is linear. The latter
becomes evident from the above expression for the surface current.
That means that, at least, linear and quadratic terms vanish after
integration of (\ref{J}) over $z$.

By convention,  the "nonequilibrium" current could be defined as a
part of (\ref{J}) which is proportional to $n_F^l-n_F^r$. In
symmetric QW this current density is an even function of $z$ and,
hence, the corresponding total nonequilibrium current is finite.
However, one can not define unambiguously the dissipative part of
the spin current using only its definition (\ref{J}). Calculation
of the spin accumulation at the sample boundary would be helpful
to clarify the physical meaning of Eq. (\ref{J}).

\subsection{Spin orientation}

In order to calculate the spin density induced by the tunneling
current, the second term in (\ref{Hso}) must be taken into account.
It causes spin flip processes upon transmission and reflection of
particles incident onto the double barrier structure. Therefore, we
label spin variables of wave functions by two indices, as
$\psi_{\alpha\beta}$, where $\alpha$ denotes the spin polarization
of the incident wave. Treating such functions as matrices, the spin
density can be expressed as
\begin{equation}\label{S}
\mathbf{S}(z)=\frac{1}{2}\sum_{\mathbf{k}_{\|},\gamma}\int_0^\infty
\frac{dk}{2\pi}Tr[\psi^{\gamma
+}\bm{\sigma}\psi^{\gamma}]n_F^{\gamma}(E)\,,
\end{equation}
where $\bm{\sigma}=(\sigma_x,\sigma_y,\sigma_z)$ is the vector of
Pauli matrices. We calculate (\ref{S}) in the first order
perturbation theory with respect to $\vartheta(z)$. Since a
commutator of two terms in (\ref{Hso}) is proportional to
$\sigma_z$, one should expect the $z$-component of $\mathbf{S}(z)$
to be finite. Further, the time inversion symmetry dictates
\begin{equation}\label{T}
\sum_{\gamma=l,r}Tr[\psi^{\gamma
+}_{\mathbf{k}_{\|}}\bm{\sigma}\psi^{\gamma}_{\mathbf{k}_{\|}}]=-\sum_{\gamma=l,r}Tr[\psi^{\gamma
+}_{-\mathbf{k}_{\|}}\bm{\sigma}\psi^{\gamma}_{-\mathbf{k}_{\|}}]\,.
\end{equation}
Applying this relation to Eq. (\ref{S}) the latter is transformed to
\begin{equation}\label{S2}
S_z(z)=\sum_{\mathbf{k}_{\|}}\int_0^\infty \frac{dk}{4\pi}
Tr[\psi^{l +}\sigma_z\psi^{l}](n_F^{l}(E)-n_F^{r}(E))\,.
\end{equation}
It immediately follows from this expression that the spin density is
zero in the equilibrium state.

The first order correction to the wave function can be written in
terms of the retarded Green function, so that
\begin{equation}\label{dpsi}
\psi^{l}_{\alpha\beta}(z)=\psi^{l}_{\alpha}(z)\delta_{\alpha\beta}+\int
dz^{\prime}G_{\beta}(z,z^{\prime})V_{\beta\alpha}(z^{\prime})\psi^{l}_{\alpha}(z^{\prime})\,,
\end{equation}
where $V_{\beta\alpha}(z^{\prime})$ is a matrix element of the
second term in (\ref{Hso}) and $\psi^{l}_{\alpha}(z)$ is the
unperturbed wave function. In its turn, the Green function is given
by
\begin{eqnarray}\label{G}
G_{\beta}(z,z^{\prime})&=&-i\frac{m_r}{kt_{\beta}}
[\psi^{l}_{\beta}(z)\psi^{r}_{\beta}(z^{\prime})\theta(z-z^{\prime})
\nonumber
\\
&+&\psi^{r}_{\beta}(z)\psi^{l}_{\beta}(z^{\prime})\theta(z^{\prime}-z)].
\end{eqnarray}
In the considered case of a symmetric heterostructure, it is easy
to see that the transmission coefficient $t_{1}=t_{2}\equiv t$. At
the same time, the reflection is spin dependent through its phase.
We note that, according to (\ref{theta}), the matrix elements
$V_{\beta\alpha}$ are proportional to $k_x^2-k_y^2$. After
integration in (\ref{S}) over angles of the vector
$\mathbf{k}_{\|}$ this expression turns to 0. It does not
necessary happen for other than [001] crystal orientations. We
will not consider such an opportunity here. Also, we will not
discuss here other than Dresselhaus's SOI effects, for example,
due to strain, or due to potential gradients along $x,y$ axes.
Instead, let us consider a situation when electrons are confined,
say, in $y$-direction. In this case $k_x^2-k_y^2$ becomes
$k_x^2-\overline{k_y^2}_n$, where the overline and the label $n$
denote averaging of the momentum operator over $n$-th quantum
eigenstate in y-direction. Thus, the symmetry between $x$ and $y$
directions is broken and $S_z$ does not turn to zero. Below we
will assume a parabolic confinement. After averaging over the
$y$-direction the spin-orbit interaction (\ref{Hso}) depends only
on $k_x$, with $k_{\|}=|k_x|$.

Substituting (\ref{G}) into (\ref{dpsi}) and then into Eq.
(\ref{S2}) we express $S_z(z)$ through unperturbed eigenstates.
The latter are calculated for a square barrier structure described
above. For an order of magnitude evaluation of the spin density,
the result can be written in an analytical form. Calculations are
strongly simplified if only resonance terms in (\ref{S2}) are
taken into account. One may also make use of the small parameter
$k_w^2/2m_w U$, where $k_w$ is the wave vector in the
$z$-direction within QW and $U$ is the barrier height. At the
lowest transmission resonance $k_w=k_0\approx \pi/2d$. By this
way, in the leading approximation the spin density in the center
of the well (z=0) and the center of the wire (y=0) can be written
as
\begin{eqnarray}\label{finS}
S_z&=-&\frac{4}{\pi}\sum_{k_x,n}|\phi_n(0)|^2\int_0^\infty
dk\frac{\Gamma^4 }{[(k_w-k_0)^2+\Gamma^2]^2}\frac{
m_r m_b}{\kappa k}  \nonumber \\
&\times & \mathcal{A}\vartheta_n (d)h^2\cosh^2\kappa
b\sinh2\kappa b \nonumber \\
&\times &  [n_F^{l}(E)-n_F^{r}(E)]\,,
\end{eqnarray}
where $h=k_{x}(\beta_w-\beta_b$, $\kappa\approx\sqrt{2m_b U}$. The
phase $\vartheta_n(z)$ is obtained from Eq. (\ref{theta}) by
averaging $k_y^2$ with oscillatory wave functions $\phi_n(y)$. The
width of the transmission resonance in $k$ space is given by
$\Gamma=(m^2_b/ m_r m_w)(k k_w/\kappa^2 d)\sinh^{-2}\kappa b$, where
$m_r, m_b, m_w$ are effective masses in reservoirs, barriers and QW,
respectively. $\mathcal{A}$ is a dimensionless function of $k$. The
value of this function is close to 1 in the range of parameters
under consideration.

For a numerical evaluation the following parameters have been
taken: $d$=100\AA, $b$=40\AA, electron density in reservoirs
$n$=10$^{23}$ m$^{-3}$, the quantization energy of the parabolic
confinement $\hbar \omega$ = 4 meV, $\gamma$=27 eV \AA$^3$.
\cite{Winkler} Other parameters correspond to the InAs quantum
well, In$_{0,53}$Ga$_{0,47}$As barrier, and In$_{0,9}$Ga$_{0,1}$As
reservoirs. With such parameters we calculated $\Gamma$= 10$^4$
cm$^{-1}$ and $\vartheta(d)\approx 10^{-3}$. From (\ref{finS}), in
the linear regime $\Delta\mu=\mu_l-\mu_r \ll E_F$,  the spin
density can be evaluated as $S_z \approx 0,1\Delta\mu$ meV$^{-1}$
$\mu$m$^{-3}$. Although the spin density is not high,
nevertheless, in the range of 1 mV terminal voltages it can be
detected by the Kerr rotation method \cite{Awschalom}. The
parameters of the heterostructure can also be optimized to reach
the higher density. Experimentally, the total spin polarization
can be enhanced in superlattices.

\subsection{Spin dependent transmittance}

The second term in (\ref{Hso}) causes spin flips of a particle
transmitting through the double barrier structure. The spin
dependent transmission is obtained from (\ref{G}) and (\ref{dpsi})
where  $\psi^l$ is calculated at $z>d+b$. By this way, near the
resonance, the spin flip transmittance becomes
\begin{eqnarray}\label{t}
\Delta t_{1,2}&=& -i16
t^2\vartheta(d)\mathcal{B}\frac{m_b^3m_r}{m_w^2}\frac{h^2k_w^2}{\kappa^3k}\frac{\cosh^3\kappa
b}{\sinh\kappa b}\,,\nonumber \\
\Delta t_{2,1}&=&-\Delta t_{1,2}
\end{eqnarray}
where $\mathcal{B}\sim 1$ is a dimensionless factor. It is easy to
see that, according to (\ref{t}), $\Delta t_{1,2}$ transforms an
unpolarized flux of electrons, say, from the left reservoir into a
flux of spin polarized electrons, with the polarization in
z-direction. Indeed, according to our choice of the spin basis,
the z-polarization is obtained by taking an average of the
$\sigma_y$ spin operator. Eq. (\ref{t}) gives rise to two spinors
corresponding to two possible initial spin polarizations:
$\psi_1=(t, \Delta t_{1,2})$ and $\psi_2=(\Delta t_{2,1},t)$. For
an unpolarized source, after summing up averages of $\sigma_y$
with these spinors, one obtains a finite result, taking into
account that near the resonance $t=i\Gamma[k_w-k_0+i\Gamma]^{-1}
\approx 1$. It should be noted, that besides (\ref{t}) $\Delta
t_{1,2}$ contains one more term which, however, does not result in
a resonant polarized transmission. This term has been neglected.
The transmission coefficient obtained from (\ref{t}) is symmetric
with respect to $x\rightarrow -x,y\rightarrow -y$ and, hence,
remains finite after the angular averaging, in contrast to a spin
filtering effect considered in Ref.\onlinecite{Voskoboynikov,
Perel}. Close to the Fermi energy and with the same numerical
parameters used for evaluation of (\ref{finS}) we obtain $\Delta
t_{1,2}\approx 3\cdot 10^{-3}$. In spite of its small value, this
coefficient can lead to a noticeable accumulation of the spin
polarization in a reservoir with small spin relaxation rate. For
example, let us take a reservoir of 1 $\mu$m$^3$ volume. The spin
polarization flux carried from the left reservoir through the wire
of the length $L$ is given by
\begin{equation}\label{Is}
I_s=\frac{L}{\pi}\sum_{k_x,n}\int_0^\infty dk \frac{k}{m_r}
\textrm{Im}[ t^*\Delta t_{1,2}][n_F^{l}(E)-n_F^{r}(E)].
\end{equation}
Taking $\mu_l-\mu_r =1$meV, $L=1\mu$m, the typical spin relaxation
time in bulk semiconductors 1 ns ,\cite{Awschalom2} and all the
rest of the parameters as in above evaluations, we get the spin
density in the reservoir around 0,5 $\mu$m$^{-3}$, which is within
the optical detection range.\cite{Awschalom}

\section{Conclusion}

We considered the spin orbit effects associated with resonant
tunneling of electrons through a symmetric double barrier structure.
Two contributions to SOI have been taken into account, namely, the
Dresselhaus SOI and the spin-orbit interaction induced by gradients
of heterostructure material parameters. We found out that the
vertical transport of electrons gives rise to the spin current
flowing parallel to heterointerfaces, as well as to the spin
polarization within QW. These effects are analogous to the spin-Hall
effect and the electric spin orientation, intensively studied
recently for 2DEG. A distinction with these traditional studies is
that instead of electron wave functions confined in QW we employ
eigenstates corresponding to resonant electron transmission.
Moreover, instead of an electric field parallel to QW, the vertical
bias has been considered. In compliance with such a 3D model, we
calculated a distribution in z-direction of the spin-Hall current
density and spin polarization. This dependence reveals interesting
structure, such as "surface" currents flowing along the
heterointerfaces. The spin-Hall current density does not turn to 0
with the external bias, thus signalling about existence of the
equilibrium spin current density, which was found to be linear with
respect to the spin-orbit interaction. At the same time, in a
symmetric double well structure the net equilibrium current,
obtained by integration of the current density over $z$, turns to 0.
Although the spin-Hall and the equilibrium currents take place in
absence of the Dresselhaus SOI, the latter is necessary to obtain
the finite out of plane spin density within QW. It was found out
that the most important is a part of the Dresselhaus SOI which is
proportional to the $z$-component of the electron momentum operator.
This SOI also gives rise to spin dependent transmission. Due to
interplay of the Dresselhaus SOI and the spin-orbit interaction
induced by gradients of heterostructure material parameters, an
unpolarized and isotropic in $x,y$ directions beam of electrons
becomes polarized in the $z$-direction after tunneling through
double barrier QW.

The effects discussed in this paper can be interesting in
application to metal surfaces with strong spin-orbit effects
associated with surface states.\cite{Ast} The STM set up with a
magnetic tip could be employed for studying spin related effects.
The above model, however,  must be modified to take into account
spin-orbit interactions typical for a particular metal surface.

This work was supported by RFBR Grant No 060216699, the National
Science Council of ROC under Grant Nos. NSC95-2112-M-009- 004,
NSC93-2119-M-007-002 (NCTS), and the MOE-ATU Grant.


\begin{thebibliography}{99}
\bibitem{theor}
H.-A. Engel, E. I. Rashba, B. I. Halperin, Handbook of Magnetism
and Advanced Magnetic Materials, Vol. 5, Wiley
(arXiv:cond-mat/0603306); J. Schliemann, Int. J. Mod. Phys. B 20,
1015 (2006).

\bibitem{exp}
J. Wunderlich, B. Kaestner, J. Sinova, and T. Jungwirth, Phys. Rev.
Lett. \textbf{94}, 047204 (2005).

\bibitem{Awschalom}
Y. K. Kato, R. C. Myers, A. C. Gossard, and D. D.
Awschalom, Science \textbf{306}, 1910 (2004)


\bibitem{Edelstein}
V.M. Edelstein, Solid State Commun., \textbf{73}, 233 (1990); F. T.
Vas'ko, N. A. Prima, Sov. Phys. Solid State, \textbf{21}, 994
(1979); E. L. Ivchenko, G. Pikus, JETP Lett. \textbf{27}, 604
(1978); L. S. Levitov, Y. N. Nazarov, G. M. Eliashberg, Sov. Phys.
JETP, \textbf{61}, 133 (1985); J. I. Inoue, G. E. W. Bauer, and L.W.
Molenkamp, Phys. Rev. B \textbf{67}, 033104 (2003)


\bibitem{Dresselhaus}
G. Dresselhaus, Phys. Rev. \textbf{100}, 580 (1955)

\bibitem{Bassani}
E. A. de Andrada e Silva, G. C. La Rocca, F. Bassani, Phys. Rev. B
\textbf{55}, 16293 (1997)
\bibitem{Winkler}
R. Winkler, Spin-Orbit Couplig Effects in Two- Dimensional
Electron and Hole Systems (Springer, Berlin, 2003).

\bibitem{RashbaSOI}
Yu. A. Bychkov and E. I. Rashba, J. Phys. C \textbf{17}, 6039 (1984)

\bibitem{Voskoboynikov}
A. Voskoboynikov, S. S. Liu, and C. P. Lee, Phys. Rev. B
\textbf{59}, 12514 (1999); E.A. de Andrada e Silva and G.C. La
Rocca, Phys. Rev. B \textbf{59}, 15583 (1999).

\bibitem{Perel}
V.I. Perel', S.A. Tarasenko, I.N. Yassievich, S. D. Ganichev, V.
V. Bel'kov, and W. Prettl, Phys. Rev. B \textbf{67}, 201304
(2003); M. M. Glazov, P. S. Alekseev, M. A. Odnoblyudov, V. M.
Chistyakov, S. A. Tarasenko, and I.N. Yassievich, Phys. Rev. B
\textbf{71}, 155313 (2005); L. G. Wang, Wen Yang, and Kai Chang,
Phys. Rev. B \textbf{72}, 153314 (2005).

\bibitem{Rashba}
E. I. Rashba, Phys. Rev. B \textbf{68}, 241315 (2003)
\bibitem{Awschalom2}
I. Malajovich, J. M. Kikkawa, and D. D. Awschalom, Phys. Rev. B
\textbf{84}, 1015 (2000)

\bibitem{Ast}
C. R. Ast, {\it et. al.}  Phys. Rev. Lett. \textbf{98}, 186807
(2007)

\end{thebibliography}
\end{document}